# Quantum Algorithm for Structure-Based Virtual Drug Screening Using Classical Force Fields

Pei-Kun Yang

E-mail: peikun@isu.edu.tw

## Keywords

binding energy, protein-ligand conformations, molecular docking, force field scoring, grid-based binding pocket

## Abstract

Structure-based virtual screening must address a combinatorial explosion arising from up to $10^{60}$ drug-like molecules, multiple conformations of proteins and ligands, and all possible spatial translations and rotations of ligands within the binding pocket. Although these calculations are inherently parallelizable, their sheer volume remains prohibitive for classical CPU/GPU resources. Quantum computing offers a promising solution: by using $n$ qubits to compute the binding energy of a single protein-ligand pair and $m$ additional qubits to encode different configurations, the algorithm can simultaneously evaluate $2^m$ combinations in a single quantum execution. To realize this potential, we propose a quantum algorithm that integrates classical force field models to compute electrostatic and van der Waals interactions on discretized grid points. Binding energy calculations are reformulated as matrix-based inner products, while ligand translations and rotations are encoded using unitary operations. This approach circumvents explicit distance calculations and provides a scalable, quantum-enhanced framework for efficient and high-dimensional binding energy estimation in drug discovery.

## Introduction

Structure-based drug screening involves calculating the binding free energy between proteins and ligands to identify potential ligands with a high affinity for the target protein [1-4]. Even when focusing solely on small, drug-like molecules, the chemical space for possible compounds is estimated to be as large as $10^{60}$ [5, 6]. Ligand libraries, such as ZINC-22, contain over $10^{10}$ purchasable ligands, any of which could demonstrate therapeutic potential [7].

Accurate computation of binding energies or scoring functions in structure-based virtual screening depends on the conformations of protein-ligand complexes [8, 9]. Typically, only the conformation of a protein in solution or bound to a specific ligand is known, along with the conformation of the ligand in either vacuum or bound states. The exact conformations of proteins complexed with all possible ligands are generally unknown. This limitation necessitates generating a range of potential conformations for both protein and ligand, covering the relevant conformational space comprehensively. By calculating binding energies across all possible conformational combinations, it becomes feasible to identify complexes with the lowest binding free energy.

In addition to conformation, determining the spatial positioning of the ligand within the binding pocket is crucial. Potential spatial configurations span three translational and three rotational dimensions, significantly expanding the search space [10-13]. This computational challenge arises from the vast number of protein-ligand conformation, translation, and rotation combinations that must be evaluated. This combinatorial problem aligns well with tensor product operations, a foundational aspect of quantum computing [14, 15]. By incorporating protein and ligand conformations, tensor product calculations enable the generation of all possible combinations within the binding site. A single quantum algorithm can then be designed to efficiently compute binding energies for these configurations, making quantum computing particularly advantageous for structure-based drug screening.

Traditional CPU/GPU-based methods for calculating binding energies generally fall into two categories: empirical scoring functions and machine learning-based approaches [10, 11, 16-18]. Empirical scoring functions use predefined mathematical formulas to approximate binding energies, while machine learning models rely on extensive datasets to predict binding affinities. Recent advancements in quantum computing—such as quantum machine learning and quantum algorithms tailored for quantum chemistry—have opened new possibilities for more accurate binding energy calculations [19-25].

This study introduces a quantum algorithm to calculate electrostatic and van der Waals energies using force fields [26, 27]. This novel approach manages the extensive range of conformations and spatial positioning required in structure-based drug screening, enabling efficient computation of binding energies across all potential configurations. Leveraging tensor product operations, the algorithm accommodates the vast combinatorial search space of protein-ligand conformations, offering rapid evaluation of binding energies that would otherwise be computationally prohibitive. This advancement illustrates the potential of quantum algorithms to transform the drug discovery process, providing a scalable and efficient solution for high-dimensional molecular interaction calculations.

## Theory

In this study, we present a quantum algorithm developed to calculate electrostatic and van der Waals potentials between proteins and ligands. The binding energy is separated into electrostatic and van der Waals components. Calculating electrostatic energy requires determining atomic distances, which involves squaring and square root operations, leading to complex quantum circuits, particularly when binary representations of real numbers are used. Given current limitations in quantum hardware, simplified circuits are advantageous. To streamline calculations, the binding pocket is discretized into grid points, allowing us to calculate the electrostatic potential generated by protein atomic charges at each point and to approximate ligand atomic contributions with grid charges [10]. The quantum computer then calculates the inner product between grid potentials and charges, an operation efficiently performed through matrix multiplication, a fundamental quantum process.

To formalize the electrostatic potential calculation, we define Equation (1) (supplementary text, note S1):

$$\boldsymbol{E}_{\text{ele}} = \Phi_{\text{grid}} \hat{\boldsymbol{\Phi}}_{\text{grid}} \boldsymbol{q}_{\text{grid}} \tag{1}$$

where $\boldsymbol{E}_{\text{ele}}$ is a vector whose first element represents the electrostatic potential energy between the protein and ligand, and $\hat{\boldsymbol{\Phi}}_{\text{grid}}$ is a unitary matrix defined in Equation (2):

$$\hat{\boldsymbol{\Phi}}_{\text{grid}}[1,:] = \begin{bmatrix} \hat{\Phi}_{\text{grid},1} & \hat{\Phi}_{\text{grid},2} & \dots \end{bmatrix} \tag{2}$$

where the first row describes the normalized electrostatic potential contributed by protein atomic charges at each grid point. Due to the unitary matrix normalization, grid potentials $\hat{\Phi}_{\text{grid},k}$ are defined as:

$$\hat{\Phi}_{\text{grid},k} = \frac{\Phi_{\text{grid},k}}{\Phi_{\text{grid}}} \tag{3}$$

where $\Phi_{\text{grid},k}$ is the electrostatic potential from protein atoms at grid point $k$, and $\Phi_{\text{grid}}$ is the root mean square of $\Phi_{\text{grid},k}$.

In Equation (1),

$$\boldsymbol{q}_{\text{grid}} = \begin{bmatrix} q_{\text{grid},1} & q_{\text{grid},2} & \cdots \end{bmatrix}^{\text{T}} \tag{4}$$

$q_{\text{grid},k}$ represents the charge at grid point $k$, approximating the contribution of ligand atomic charges through grid charges.

For van der Waals potential energy, we calculate the interaction potentials between the protein and each ligand atomic type $t$. The van der Waals potential energy for ligand atomic type $t$ is represented in Equation (5) (supplementary text, note S2):

$$\boldsymbol{E}^{t}_{\text{vdW}} = E^{t}_{\text{vdW,grid}} \hat{\boldsymbol{E}}^{t}_{\text{vdW,grid}} \boldsymbol{N}^{t}_{\text{grid}} \tag{5}$$

where $\boldsymbol{E}^{t}_{\text{vdW}}$ is a vector, with its first element corresponding to the van der Waals potential energy between protein and ligand atoms of type $t$. For each atomic type, a unitary matrix $\hat{\boldsymbol{E}}^{t}_{\text{vdW,grid}}$ is defined in Equation (6):

$$\hat{\boldsymbol{E}}^{t}_{\text{vdW,grid}}[1,:] = \begin{bmatrix} \hat{E}^{t}_{\text{vdW,grid},1} & \hat{E}^{t}_{\text{vdW,grid},2} & \end{bmatrix} \tag{6}$$

where the first row represents the normalized van der Waals potential energy from protein atoms at each grid point of type $t$:

$$\hat{E}^{t}_{\text{vdW,grid},k} = \frac{E^{t}_{\text{vdW,grid},k}}{E^{t}_{\text{vdW,grid}}} \tag{7}$$

where $E^{t}_{\text{vdW,grid},k}$ is the van der Waals potential energy from protein atoms at grid point $k$ of type $t$, and $E^{t}_{\text{vdW,grid}}$ is the root mean square of $E^{t}_{\text{vdW,grid},k}$.

In Equation (5),

$$\boldsymbol{N}^{t}_{\text{grid}} = \begin{bmatrix} N^{t}_{\text{grid},1} & N^{t}_{\text{grid},2} & \cdots \end{bmatrix}^{\text{T}} \tag{8}$$

The occupancy rate of ligand atoms of type $t$ at grid point $k$ is denoted by $N^{t}_{\text{grid},k}$.

The total binding energy, combining both electrostatic and van der Waals energies, is represented by Equation (9), which integrates Equations (1) and (5):

$$\boldsymbol{E} = \left(2^{nt/2} L^{\text{type}}\right)\left(\text{H}^{\otimes nt} \otimes \text{I}^{\otimes ng}\right) \mathbf{U}_{\text{grid}} \left(\frac{\mathbf{O}_{\text{grid}}}{L^{\text{type}}}\right) \tag{9}$$

This calculation requires the construction of a block diagonal unitary matrix $\mathbf{U}_{\mathrm{grid}}$, which integrates both the electrostatic potential $\hat{\mathbf{\Phi}}_{\mathrm{grid}}$ and the van der Waals matrix $\hat{\boldsymbol{E}}^{t}_{\mathrm{vdW,grid}}$ across various atomic types.

$$\mathbf{U}_{\mathrm{grid}} = \mathrm{diag}\left(\hat{\mathbf{\Phi}}_{\mathrm{grid}}, \hat{\boldsymbol{E}}^{1}_{\mathrm{vdW,grid}}, \hat{\boldsymbol{E}}^{2}_{\mathrm{vdW,grid}}, \ldots\right) \tag{10}$$

Here, diag(...) represents a diagonal matrix.

In Equation (9), $\mathbf{O}_{\mathrm{grid}}$ combines $\Phi_{\mathrm{grid}}$ and $\boldsymbol{q}_{\mathrm{grid}}$ from Equation (1) with $E^{t}_{\mathrm{vdW,grid}}$ and $\boldsymbol{N}^{t}_{\mathrm{grid}}$ from Equation (5):

$$\mathbf{O}_{\mathrm{grid}} = \left[\Phi_{\mathrm{grid}}\boldsymbol{q}_{\mathrm{grid}} \quad E^{1}_{\mathrm{vdW,grid}}\boldsymbol{N}^{1}_{\mathrm{grid}} \quad E^{2}_{\mathrm{vdW,grid}}\boldsymbol{N}^{2}_{\mathrm{grid}} \quad \ldots\right]^{\mathrm{T}} \tag{11}$$

To normalize, we calculate the root mean square to obtain a length $L^{\mathrm{type}}$, ensuring that $\mathbf{O}_{\mathrm{grid}}/L^{\mathrm{type}}$ represents a unit vector.

In Equation (9), $(\mathrm{H}^{\otimes nt} \otimes \mathrm{I}^{\otimes ng})$ is used to sum the electrostatic and van der Waals energies for different ligand atom types. Here, H is the Hadamard gate, and I is the Identity gate. The value $2^{nt}$ must exceed the combined total of electrostatic and ligand atom types, and $2^{ng}$ must exceed the number of grid points. The matrix $\boldsymbol{E} = [E, \ldots]^{\mathrm{T}}$ contains the potential energy terms, with the first element $E$ representing the interaction potential energy between the protein and ligand.

The term $(\mathrm{H}^{\otimes nt} \otimes \mathrm{I}^{\otimes ng})\mathbf{U}_{\mathrm{grid}}(\mathbf{O}_{\mathrm{grid}}/L^{\mathrm{type}})$ in Equation (9) can be computed on a quantum computer, and the probability from the measurement is given by:

$$p=\left(\left(\mathrm{H}^{\otimes nt} \otimes \mathrm{I}^{\otimes ng}\right)\mathbf{U}_{\mathrm{grid}}\mathbf{O}_{\mathrm{grid}} / L^{\mathrm{type}}\right)^{2} \tag{12}$$

The probability $p_0$ in the $|0>^{\otimes(nt+ng)}$ quantum state represents the potential energy. When substituting into Equation (9), the potential energy between the protein and ligand becomes:

$$E=\sqrt{p_0}\left(2^{nt/2} L^{\mathrm{type}}\right) \tag{13}$$

Since binding energy is generally negative, the probability output corresponds to the square of the quantum state. To ensure a non-negative output, a grid point outside the atom region can be set, such as $\Phi_{\mathrm{grid,last}} = c$ and $q_{\mathrm{grid,last}} = 1$, where $c$ is a positive constant. After calculating the binding energy, $c$ can be subtracted as needed.

This quantum algorithm also calculates binding energies across multiple protein and ligand conformational combinations. Equation (9) provides the binding energy for a single conformation. For protein conformations $\leq 2^{nrc}$ and ligand conformations $\leq 2^{nlc}$, the combined binding energy across all configurations is calculated as follows:

$$\boldsymbol{E} = \left( 2^{\frac{nt+nrc}{2}} L^{\text{con}} \right) \left( \mathrm{I}^{\otimes nrc+nlc} \otimes \mathrm{H}^{\otimes nt} \otimes \mathrm{I}^{\otimes ng} \right) \cdot \mathrm{diag}\left( \mathrm{I}^{\otimes nlc} \otimes \mathbf{U}^{1}_{\text{grid}}, \mathrm{I}^{\otimes nlc} \otimes \mathbf{U}^{2}_{\text{grid}}, \ldots \right) \left( \mathrm{H}^{\otimes nrc} \otimes \frac{1}{L^{\text{con}}} \left[ \mathbf{O}^{1}_{\text{grid}} \quad \mathbf{O}^{2}_{\text{grid}} \quad \ldots \right]^{\mathrm{T}} \right) \tag{14}$$

$\mathbf{U}^{i}_{\text{grid}}$ represents the normalized electrostatic and van der Waals potentials for the $i^{\text{th}}$ protein conformation at each grid point. To ensure compatibility with each ligand conformation, $\mathrm{I}^{\otimes nlc}$ is used to generate $2^{nlc}$ instances of $\mathbf{U}^{i}_{\text{grid}}$. Each ($\mathrm{I}^{\otimes nlc} \otimes \mathbf{U}^{i}_{\text{grid}}$) is positioned along the diagonal of a block diagonal unitary matrix. $\mathbf{O}^{j}_{\text{grid}}$ represents the weighted charges and atomic occupancy rates for the $j^{\text{th}}$ ligand conformation. Calculating the length of $[\mathbf{O}^{1}_{\text{grid}}, \mathbf{O}^{2}_{\text{grid}}, \ldots]^{\mathrm{T}}$ provides $L^{\text{con}}$, which is then normalized. The Hadamard operator $\mathrm{H}^{\otimes nrc}$ generates $2^{nrc}$ sets. On the right-hand side of Equation (14), $2^{(nt+nrc)/2} L^{\text{con}}$ is a constant, and the subsequent matrix products can be computed on a quantum computer, yielding $2^{nrc+nlc}$ binding energy combinations in a single execution.

Furthermore, ligand translations along the $x$, $y$, and $z$ axes are represented by unitary matrices, such as $T_x^g$, $T_y^g$, and $T_z^g$, which shift by $g$ grid points along each axis (supplementary text, note S3).

The distribution $\mathbf{O}^{\text{trans}}_{\text{grid}}$, which represents various translational combinations of $\mathbf{O}_{\text{grid}}$ along the $x$, $y$, and $z$ axes, is obtained using the following equation:

$$\mathbf{O}^{\text{trans}}_{\text{grid}} = 2^{\frac{ntx+nty+ntz}{2}} \tilde{\boldsymbol{T}}_z \tilde{\boldsymbol{T}}_y \tilde{\boldsymbol{T}}_x \left( \mathrm{H}^{\otimes (ntx+nty+ntz)} \otimes \mathbf{O}_{\text{grid}} \right) \tag{15}$$

In Equation (15)

$$\tilde{\boldsymbol{T}}_x = \mathrm{I}^{\otimes (nty+ntz)} \otimes \mathrm{diag}\left( \mathrm{I}^{\otimes nt} \otimes T_x^1, \mathrm{I}^{\otimes nt} \otimes T_x^2, \ldots \right) \tag{16}$$

$$\tilde{\boldsymbol{T}}_y = \mathrm{I}^{\otimes ntz} \cdot \mathrm{diag}\left( \mathrm{I}^{\otimes (nt+ntx)} \otimes T_y^1, \mathrm{I}^{\otimes (nt+ntx)} \otimes T_y^2, \ldots \right) \tag{17}$$

and,

$$\tilde{\boldsymbol{T}}_z = \mathrm{diag}\left( \mathrm{I}^{\otimes (nt+ntx+nty)} \otimes T_z^1, \mathrm{I}^{\otimes (nt+ntx+nty)} \otimes T_z^2, \ldots \right) \tag{18}$$

The required qubit number for translations along each axis is limited to $\leq 2^{ntx}$, $2^{nty}$, and

$2^{ntz}$.

For ligand rotations around the $x$, $y$, and $z$ axes, changes in grid point positions are represented by unitary matrices $R_x{}^g$, $R_y{}^g$, and $R_z{}^g$ (supplementary text, note S4).

The distribution $\mathbf{O}^{\text{rot}}{}_{\text{grid}}$, representing various rotational combinations of $\mathbf{O}_{\text{grid}}$ around the $x$, $y$, and $z$ axes, can be obtained as:

$$\mathbf{O}_{\text{grid}}^{\text{rot}} = 2^{\frac{nrx+nry+nrz}{2}} \tilde{\boldsymbol{R}}_z \tilde{\boldsymbol{R}}_y \tilde{\boldsymbol{R}}_x \left( \mathrm{H}^{\otimes(nrx+nry+nrz)} \otimes \mathbf{O}_{\text{grid}} \right) \tag{19}$$

In Equation (19)

$$\tilde{\boldsymbol{R}}_x = \mathrm{I}^{\otimes(nry+nrz)} \otimes \mathrm{diag}\left( \mathrm{I}^{\otimes nt} \otimes R_x^1, \mathrm{I}^{\otimes nt} \otimes R_x^2, \ldots \right) \tag{20}$$

$$\tilde{\boldsymbol{R}}_y = \mathrm{I}^{\otimes nrz} \cdot \mathrm{diag}\left( \mathrm{I}^{\otimes(nt+nrx)} \otimes R_y^1, \mathrm{I}^{\otimes(nt+nrx)} \otimes R_y^2, \ldots \right) \tag{21}$$

and,

$$\tilde{\boldsymbol{R}}_z = \mathrm{diag}\left( \mathrm{I}^{\otimes(nt+nrx+nry)} \otimes R_z^1, \mathrm{I}^{\otimes(nt+nrx+nry)} \otimes R_z^2, \ldots \right) \tag{22}$$

The number of rotation angles around the $x$, $y$, and $z$ axes is limited to $\leq 2^{nrx}$, $\leq 2^{nry}$, and $\leq 2^{nrz}$, respectively.

The construction of $\hat{\Phi}_{\text{grid}}$ in Equation (2) and $\hat{E}_{\text{vdW,grid}}^{t}$ in Equation (6) requires generating unitary matrices with a specified first row. This is achieved using a transposed circuit structure (Figure 1) [28]. As the matrices $(ab)^{\text{T}} = b^{\text{T}}a^{\text{T}}$, the circuit layers are reversed compared to conventional quantum state preparation, with the RY gates' phase values inverted in each layer. The implementation of this method is provided in an open-source Python codebase, available at https://github.com/peikunyang/12_quantum_state.

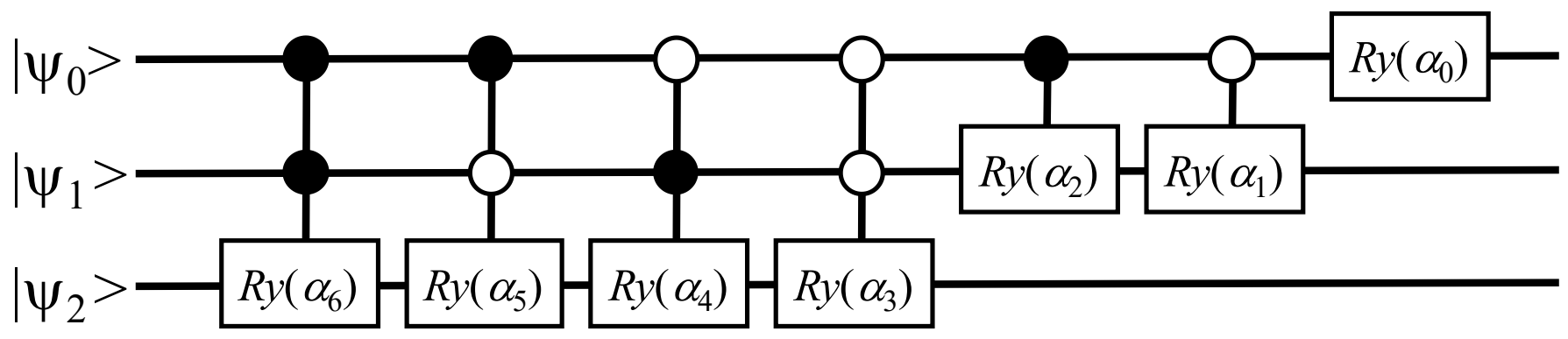

Fig. 1. Generation of a unitary matrix with specified values in the first row.

Finally, to simulate the solvent effect, a distance-dependent dielectric constant is introduced for electrostatic screening in aqueous environments. This can be directly incorporated in Equation (2) by applying the dielectric constant or, more accurately, by solving the Poisson-Boltzmann equation on a quantum computer [29].

## Conclusion

We developed a quantum algorithm for structure-based virtual drug screening, offering a novel approach to tackle the complex problem of binding energy calculations across numerous protein and ligand conformations. By employing tensor products, this algorithm efficiently manages a wide range of protein conformations, ligand types, spatial translations, and rotations, thereby addressing the challenges associated with calculating binding energies over extensive configurational spaces. In each binding energy calculation, grid points within the binding pocket are utilized, allowing us to avoid quantum algorithms that directly compute interatomic distances, which can be computationally intensive.

This quantum algorithm demonstrates the potential of quantum computing to significantly streamline and enhance the virtual drug screening process, enabling rapid and scalable solutions for high-dimensional molecular interactions. With this approach, drug discovery processes can potentially be transformed, offering a promising avenue for tackling the vast configuration space inherent in protein-ligand interactions.

## Reference

[1] Bajorath J 2002 Integration of virtual and high-throughput screening *Nature Reviews Drug Discovery* **1** 882-94

[2] Lavecchia A and Di Giovanni C 2013 Virtual screening strategies in drug discovery: a critical review *Current medicinal chemistry* **20** 2839-60

[3] Maia E H B, Assis L C, De Oliveira T A, Da Silva A M and Taranto A G 2020 Structure-based virtual screening: from classical to artificial intelligence *Frontiers in chemistry* **8** 343

[4] Ballante F, Kooistra A J, Kampen S, de Graaf C and Carlsson J 2021 Structure-based virtual screening for ligands of G protein–coupled receptors: what can

molecular docking do for you? *Pharmacological Reviews* **73** 1698-736
[5] Reymond J-L and Awale M 2012 Exploring chemical space for drug discovery using the chemical universe database *ACS chemical neuroscience* **3** 649-57
[6] Polishchuk P G, Madzhidov T I and Varnek A 2013 Estimation of the size of drug-like chemical space based on GDB-17 data *Journal of computer-aided molecular design* **27** 675-9
[7] Tingle B I, Tang K G, Castanon M, Gutierrez J J, Khurelbaatar M, Dandarchuluun C, Moroz Y S and Irwin J J 2023 ZINC-22— A free multi-billion-scale database of tangible compounds for ligand discovery *Journal of chemical information and modeling* **63** 1166-76
[8] Frimurer T M, Peters G H, Iversen L F, Andersen H S, Møller N P H and Olsen O H 2003 Ligand-induced conformational changes: improved predictions of ligand binding conformations and affinities *Biophysical journal* **84** 2273-81
[9] Ahmad E, Rabbani G, Zaidi N, Khan M A, Qadeer A, Ishtikhar M, Singh S and Khan R H 2013 Revisiting ligand-induced conformational changes in proteins: essence, advancements, implications and future challenges *Journal of Biomolecular Structure and Dynamics* **31** 630-48
[10] Goodsell D S, Sanner M F, Olson A J and Forli S 2021 The AutoDock suite at 30 *Protein Science* **30** 31-43
[11] Trott O and Olson A J 2010 AutoDock Vina: improving the speed and accuracy of docking with a new scoring function, efficient optimization, and multithreading *Journal of computational chemistry* **31** 455-61
[12] Gray J J, Moughon S, Wang C, Schueler-Furman O, Kuhlman B, Rohl C A and Baker D 2003 Protein–protein docking with simultaneous optimization of rigid-body displacement and side-chain conformations *Journal of molecular biology* **331** 281-99
[13] Pagadala N S, Syed K and Tuszynski J 2017 Software for molecular docking: a review *Biophys Rev* **9** 91-102
[14] de Lima Marquezino F, Portugal R and Lavor C 2019 *A primer on quantum computing*: Springer)
[15] Loceff M 2015 *A course in quantum computing (for the community college). Foothill College*
[16] Yang C and Zhang Y 2021 Lin_F9: A Linear Empirical Scoring Function for Protein–Ligand Docking *Journal of Chemical Information and Modeling*

[17] McNutt A T, Francoeur P, Aggarwal R, Masuda T, Meli R, Ragoza M, Sunseri J and Koes D R 2021 GNINA 1.0: molecular docking with deep learning *Journal of cheminformatics* **13** 1-20

[18] Li H, Sze K H, Lu G and Ballester P J 2021 Machine-learning scoring functions for structure-based virtual screening *Wiley Interdisciplinary Reviews: Computational Molecular Science* **11** e1478

[19] Zeguendry A, Jarir Z and Quafafou M 2023 Quantum machine learning: A review and case studies *Entropy* **25** 287

[20] Peral-García D, Cruz-Benito J and García-Peñalvo F J 2024 Systematic literature review: Quantum machine learning and its applications *Computer Science Review* **51** 100619

[21] Ma H, Liu J, Shang H, Fan Y, Li Z and Yang J 2023 Multiscale quantum algorithms for quantum chemistry *Chemical Science* **14** 3190-205

[22] Cao Y, Romero J, Olson J P, Degroote M, Johnson P D, Kieferová M, Kivlichan I D, Menke T, Peropadre B and Sawaya N P 2019 Quantum chemistry in the age of quantum computing *Chemical reviews* **119** 10856-915

[23] Aspuru-Guzik A, Dutoi A D, Love P J and Head-Gordon M 2005 Simulated quantum computation of molecular energies *Science* **309** 1704-7

[24] Daskin A and Kais S 2011 Decomposition of unitary matrices for finding quantum circuits: application to molecular Hamiltonians *The Journal of chemical physics* **134**

[25] Schuld M and Killoran N 2019 Quantum machine learning in feature hilbert spaces *Physical review letters* **122** 040504

[26] Brooks B R, Brooks III C L, Mackerell Jr A D, Nilsson L, Petrella R J, Roux B, Won Y, Archontis G, Bartels C and Boresch S 2009 CHARMM: the biomolecular simulation program *Journal of computational chemistry* **30** 1545-614

[27] Lee J, Cheng X, Jo S, MacKerell A D, Klauda J B and Im W 2016 CHARMM-GUI input generator for NAMD, GROMACS, AMBER, OpenMM, and CHARMM/OpenMM simulations using the CHARMM36 additive force field *Biophysical journal* **110** 641a

[28] Araujo I F, Park D K, Petruccione F and da Silva A J 2021 A divide-and-conquer algorithm for quantum state preparation *Scientific reports* **11** 6329

[29] Daribayev B, Mukhanbet A and Imankulov T 2023 Implementation of the hhl algorithm for solving the poisson equation on quantum simulators *Applied*

*Sciences* **13** 11491

# Quantum Algorithm for Structure-Based Virtual Drug Screening

Pei-Kun Yang

E-mail: peikun@isu.edu.tw

## Supplementary S1: Electrostatic Potential Energy Calculation

This section describes the quantum algorithm for calculating the electrostatic potential energy between a protein and a ligand. The electrostatic potential energy $E_{\text{ele}}$ is given by:

$$4\pi\varepsilon_0 E_{\text{ele}} = \sum_i \sum_j \frac{Q_i q_j}{\left|\boldsymbol{r}_i - \boldsymbol{r}_j\right|} \tag{S1.1}$$

where $Q_i$ is the charge of the $i^{\text{th}}$ protein atom, $q_j$ is the charge of the $j^{\text{th}}$ ligand atom, and $\boldsymbol{r}_i$ and $\boldsymbol{r}_j$ are their spatial coordinates.

The electrostatic potential $\Phi_{\text{grid},k}$ at grid point $k$, due to the charges on protein atoms, is expressed as:

$$4\pi\varepsilon_0 \Phi_{\text{grid},k} = \sum_i \frac{Q_i}{\left|\boldsymbol{r}_i - \boldsymbol{r}_k\right|} \tag{S1.2}$$

where $\boldsymbol{r}_k$ denotes the position of the grid point. Using linear interpolation, the charge distribution $q_{\text{grid}}$ at the grid points is determined to approximate the electrostatic effects of ligand atom charges $q_j$.

The electrostatic potential energy is then approximated as:

$$E_{\text{ele}} \approx \sum_k q_{\text{grid},k} \Phi_{\text{grid},k} \tag{S1.3}$$

For quantum computation, Equation (S1.3) is reformulated as a matrix product:

$$
\begin{bmatrix} E_{\text{ele}} \\ \\ \\ \end{bmatrix} = \begin{bmatrix} \Phi_{\text{grid},1} & \Phi_{\text{grid},2} & \cdots \\ & & \\ & & \end{bmatrix} \begin{bmatrix} q_{\text{grid},1} \\ q_{\text{grid},2} \\ \cdots \end{bmatrix} \tag{S1.4}
$$

The first element on the left-hand side corresponds to the electrostatic potential energy. Because quantum circuits are restricted to generating unitary matrices, each row of the matrix must be normalized. Therefore, the norm of the first row in the matrix from Equation (S1.4) is calculated as:

$$
\Phi_{\text{grid}} = \sqrt{\sum_k \left(\Phi_{\text{grid},k}\right)^2} \tag{S1.5}
$$

Defining $\hat{\Phi}_{\text{grid}}$ as the normalized potential:

$$
\hat{\Phi}_{\text{grid},k} = \frac{\Phi_{\text{grid},k}}{\Phi_{\text{grid}}} \tag{S1.6}
$$

so that Equation (S1.4) can be rewritten as:

$$
\begin{bmatrix} E_{\text{ele}} \\ \cdots \\ \cdots \end{bmatrix} = \Phi_{\text{grid}} \begin{bmatrix} \hat{\Phi}_{\text{grid},1} & \hat{\Phi}_{\text{grid},2} & \cdots \\ \cdots & \cdots & \cdots \\ \cdots & \cdots & \cdots \end{bmatrix} \begin{bmatrix} q_{\text{grid},1} \\ q_{\text{grid},2} \\ \cdots \end{bmatrix} \tag{S1.7}
$$

This reformulation enables the quantum circuit to construct the first matrix on the right side of the equation. In a concise form, Equation (S1.7) is expressed as:

$$
\boldsymbol{E}_{\text{ele}} = \Phi_{\text{grid}} \hat{\boldsymbol{\Phi}}_{\text{grid}} \boldsymbol{q}_{\text{grid}} \tag{1}
$$

with:

$$
\boldsymbol{E}_{\text{ele}} = \begin{bmatrix} E_{\text{ele}} & \cdots & \cdots \end{bmatrix}^{\text{T}} \tag{2}
$$

$$
\hat{\boldsymbol{\Phi}}_{\text{grid}} = \begin{bmatrix} \hat{\Phi}_{\text{grid},1} & \hat{\Phi}_{\text{grid},2} & \cdots \\ \cdots & \cdots & \cdots \end{bmatrix} \tag{3}
$$

and

$$
\boldsymbol{q}_{\text{grid}} = \begin{bmatrix} q_{\text{grid},1} & q_{\text{grid},2} & \cdots \end{bmatrix}^{\text{T}} \tag{4}
$$

## Supplementary S2: Van der Waals Potential Energy Calculation

This section presents the quantum algorithm for calculating the van der Waals potential energy $E_{\mathrm{vdW}}$ between atoms of a protein and a ligand. The van der Waals potential energy $E_{\mathrm{vdW}}$ is expressed as:

$$E_{\mathrm{vdW}} = \sum_i \sum_j \sqrt{\varepsilon_i \varepsilon_j} \left[ \left( \frac{R_{\min,i} + R_{\min,j}}{|\boldsymbol{r}_i - \boldsymbol{r}_j|} \right)^{12} - 2 \left( \frac{R_{\min,i} + R_{\min,j}}{|\boldsymbol{r}_i - \boldsymbol{r}_j|} \right)^{6} \right] \tag{S2.1}$$

where $\boldsymbol{r}_i$ and $\boldsymbol{r}_j$ are the spatial coordinates of the protein and ligand atoms, respectively, and $R_{\min}$ and $\varepsilon$ represent the van der Waals distance and energy parameters for the atoms. Similar to the electrostatic potential energy calculation, the van der Waals potential energy is evaluated between the protein and grid points, and subsequently weighted by the occupancy of ligand atoms at each grid point.

The van der Waals potential energy between a protein and grid point $k$ for ligand atom type $t$ is given by:

$$E^t_{\mathrm{vdW,grid},k} = \sum_i \sqrt{\varepsilon_i \varepsilon_t} \left[ \left( \frac{R_{\min,i} + R_{\min,t}}{|\boldsymbol{r}_i - \boldsymbol{r}_k|} \right)^{12} - 2 \left( \frac{R_{\min,i} + R_{\min,t}}{|\boldsymbol{r}_i - \boldsymbol{r}_k|} \right)^{6} \right] \tag{S2.2}$$

The occupancy $N_{\mathrm{grid},k}{}^t$ of ligand atoms of type $t$ at grid point $k$ can be estimated through linear interpolation.

Thus, the van der Waals potential energy between the protein and ligand for atom type $t$ becomes:

$$E^t_{\mathrm{vdW}} = \sum_k E^t_{\mathrm{vdW,grid},k} N^t_{\mathrm{grid},k} \tag{S2.3}$$

For quantum computation, Equation (S2.3) is reformulated as a matrix product:

$$\begin{bmatrix} E^t_{\mathrm{vdW}} \\ \\ \\ \end{bmatrix} = \begin{bmatrix} E^t_{\mathrm{vdW,grid},1} & E^t_{\mathrm{vdW,grid},2} & \cdots \\ & & \\ & & \end{bmatrix} \begin{bmatrix} N^t_{\mathrm{grid},1} \\ N^t_{\mathrm{grid},2} \\ \cdots \end{bmatrix} \tag{S2.4}$$

The first element on the left side represents the van der Waals potential energy between the protein and the ligand for atom type $t$. Because quantum circuits are constrained to generate unitary matrices, each row of the matrix must be normalized. Consequently, the norm of the first row in the first matrix on the right side of Equation (S2.4) is calculated as:

$$E^{t}_{\mathrm{vdW,grid}} = \sqrt{\sum_{k}\left(E^{t}_{\mathrm{vdW,grid},k}\right)^{2}} \tag{S2.5}$$

We define the normalized potential as:

$$\hat{E}^{t}_{\mathrm{vdW,grid},k} = \frac{E^{t}_{\mathrm{vdW,grid},k}}{E^{t}_{\mathrm{vdW,grid}}} \tag{S2.6}$$

Thus, Equation (S2.4) can be reformulated as:

$$\begin{bmatrix} E^{t}_{\mathrm{vdW}} \\ \cdots \\ \cdots \end{bmatrix} = E^{t}_{\mathrm{vdW,grid}} \begin{bmatrix} \hat{E}^{t}_{\mathrm{vdW,grid,1}} & \hat{E}^{t}_{\mathrm{vdW,grid,2}} & \cdots\cdot \\ \cdots & \cdots & \cdots \\ \cdots & \cdots & \cdots \end{bmatrix} \begin{bmatrix} N^{t}_{\mathrm{grid,1}} \\ N^{t}_{\mathrm{grid,2}} \\ \cdots \end{bmatrix} \tag{S2.7}$$

This reformulation enables the quantum circuit to construct the first matrix on the right side of the equation. In a concise form, Equation (S2.7) becomes:

$$\boldsymbol{E}^{t}_{\mathrm{vdW}} = E^{t}_{\mathrm{vdW,grid}} \hat{\boldsymbol{E}}^{t}_{\mathrm{vdW,grid}} \boldsymbol{N}^{t}_{\mathrm{grid}} \tag{5}$$

where:

$$\boldsymbol{E}^{t}_{\mathrm{vdW}} = \left[ E^{t}_{\mathrm{vdW}} \quad \cdots \quad \cdots \right]^{\mathrm{T}} \tag{S2.8}$$

$$\hat{\boldsymbol{E}}^{t}_{\mathrm{vdW,grid}} = \begin{bmatrix} \hat{E}^{t}_{\mathrm{vdW,grid,1}} & \hat{E}^{t}_{\mathrm{vdW,grid,2}} & \cdots \\ \cdots & \cdots & \cdots \end{bmatrix} \tag{6}$$

and

$$\boldsymbol{N}^{t}_{\mathrm{grid}} = \left[ \boldsymbol{N}^{t}_{\mathrm{grid,1}} \quad \boldsymbol{N}^{t}_{\mathrm{grid,2}} \quad \cdots \right]^{\mathrm{T}} \tag{8}$$

## Supplementary S3: Ligand Translation

This section illustrates ligand translation using a 2D example. The first two qubits encode grid point variations along the $y$-axis, while the next two encode variations along the $x$-axis. The quantum states for specific grid positions are represented as:

$$\begin{bmatrix} 0000 & 0001 & 0010 & 0011 \\ 0100 & 0101 & 0110 & 0111 \\ 1000 & 1001 & 1010 & 1011 \\ 1100 & 1101 & 1110 & 1111 \end{bmatrix} \tag{S3.1}$$

When the grid points are shifted by one unit to the right, the corresponding quantum states become:

$$\begin{bmatrix} 0011 & 0000 & 0001 & 0010 \\ 0111 & 0100 & 0101 & 0110 \\ 1011 & 1000 & 1001 & 1010 \\ 1111 & 1100 & 1101 & 1110 \end{bmatrix} \tag{S3.2}$$

To avoid unintended wrap-around effects in the transformed state, the grid coverage must extend beyond the occupied ligand space by at least the translation distance. The transformation matrix $T_x^1$ converts $|\psi_{\text{grid}}>$ to $|\psi'_{\text{grid}}>$:

$$T_x^1 = \begin{bmatrix} 0&0&0&1&0&0&0&0&0&0&0&0&0&0&0&0 \\ 1&0&0&0&0&0&0&0&0&0&0&0&0&0&0&0 \\ 0&1&0&0&0&0&0&0&0&0&0&0&0&0&0&0 \\ 0&0&1&0&0&0&0&0&0&0&0&0&0&0&0&0 \\ 0&0&0&0&0&0&0&1&0&0&0&0&0&0&0&0 \\ 0&0&0&0&1&0&0&0&0&0&0&0&0&0&0&0 \\ 0&0&0&0&0&1&0&0&0&0&0&0&0&0&0&0 \\ 0&0&0&0&0&0&1&0&0&0&0&0&0&0&0&0 \\ 0&0&0&0&0&0&0&0&0&0&0&1&0&0&0&0 \\ 0&0&0&0&0&0&0&0&1&0&0&0&0&0&0&0 \\ 0&0&0&0&0&0&0&0&0&1&0&0&0&0&0&0 \\ 0&0&0&0&0&0&0&0&0&0&1&0&0&0&0&0 \\ 0&0&0&0&0&0&0&0&0&0&0&0&0&0&0&1 \\ 0&0&0&0&0&0&0&0&0&0&0&0&1&0&0&0 \\ 0&0&0&0&0&0&0&0&0&0&0&0&0&1&0&0 \\ 0&0&0&0&0&0&0&0&0&0&0&0&0&0&1&0 \end{bmatrix} \tag{S3.3}$$

Similarly, a downward shift is represented as:

$$\begin{bmatrix} 1100 & 1101 & 1110 & 1111 \\ 0000 & 0001 & 0010 & 0011 \\ 0100 & 0101 & 0110 & 0111 \\ 1000 & 1001 & 1010 & 1011 \end{bmatrix} \tag{S3.4}$$

The transformation matrix $T_y{}^1$ is defined to achieve this shift:

$$T_y^1 = \begin{bmatrix}
0&0&0&0&0&0&0&0&0&0&0&0&1&0&0&0 \\
0&0&0&0&0&0&0&0&0&0&0&0&0&1&0&0 \\
0&0&0&0&0&0&0&0&0&0&0&0&0&0&1&0 \\
0&0&0&0&0&0&0&0&0&0&0&0&0&0&0&1 \\
1&0&0&0&0&0&0&0&0&0&0&0&0&0&0&0 \\
0&1&0&0&0&0&0&0&0&0&0&0&0&0&0&0 \\
0&0&1&0&0&0&0&0&0&0&0&0&0&0&0&0 \\
0&0&0&1&0&0&0&0&0&0&0&0&0&0&0&0 \\
0&0&0&0&1&0&0&0&0&0&0&0&0&0&0&0 \\
0&0&0&0&0&1&0&0&0&0&0&0&0&0&0&0 \\
0&0&0&0&0&0&1&0&0&0&0&0&0&0&0&0 \\
0&0&0&0&0&0&0&1&0&0&0&0&0&0&0&0 \\
0&0&0&0&0&0&0&0&1&0&0&0&0&0&0&0 \\
0&0&0&0&0&0&0&0&0&1&0&0&0&0&0&0 \\
0&0&0&0&0&0&0&0&0&0&1&0&0&0&0&0 \\
0&0&0&0&0&0&0&0&0&0&0&1&0&0&0&0
\end{bmatrix} \tag{S3.5}$$

## Supplementary S4: Ligand Rotation

In this section, ligand rotation is illustrated using a 2D example. The first two qubits encode grid point variations along the $y$-axis, and the next two encode variations along the $x$-axis. The quantum state for a specific grid point is given by:

$$\begin{bmatrix} 0000 & 0001 & 0010 & 0011 \\ 0100 & 0101 & 0110 & 0111 \\ 1000 & 1001 & 1010 & 1011 \\ 1100 & 1101 & 1110 & 1111 \end{bmatrix} \tag{S4.1}$$

When the grid points are rotated by 90 degrees clockwise around the z-axis, the state transforms into:

$$\begin{bmatrix} 1100 & 1000 & 0100 & 0000 \\ 1101 & 1001 & 0101 & 0001 \\ 1110 & 1010 & 0110 & 0010 \\ 1111 & 1011 & 0111 & 0011 \end{bmatrix} \tag{S4.2}$$

The transformation matrix $R_z{}^{90}$ converts $|\psi_{\mathrm{grid}}>$ to $|\psi'_{\mathrm{grid}}>$:

$$R_x^{90} = \begin{bmatrix} 0&0&0&1&0&0&0&0&0&0&0&0&0&0&0&0 \\ 0&0&0&0&0&0&0&1&0&0&0&0&0&0&0&0 \\ 0&0&0&0&0&0&0&0&0&0&0&1&0&0&0&0 \\ 0&0&0&0&0&0&0&0&0&0&0&0&0&0&0&1 \\ 0&0&1&0&0&0&0&0&0&0&0&0&0&0&0&0 \\ 0&0&0&0&0&0&1&0&0&0&0&0&0&0&0&0 \\ 0&0&0&0&0&0&0&0&0&0&1&0&0&0&0&0 \\ 0&0&0&0&0&0&0&0&0&0&0&0&0&0&1&0 \\ 0&1&0&0&0&0&0&0&0&0&0&0&0&0&0&0 \\ 0&0&0&0&0&1&0&0&0&0&0&0&0&0&0&0 \\ 0&0&0&0&0&0&0&0&0&1&0&0&0&0&0&0 \\ 0&0&0&0&0&0&0&0&0&0&0&0&0&1&0&0 \\ 1&0&0&0&0&0&0&0&0&0&0&0&0&0&0&0 \\ 0&0&0&0&1&0&0&0&0&0&0&0&0&0&0&0 \\ 0&0&0&0&0&0&0&0&1&0&0&0&0&0&0&0 \\ 0&0&0&0&0&0&0&0&0&0&0&0&1&0&0&0 \end{bmatrix} \tag{S4.3}$$